\documentclass[doublecol]{epl2} 
\usepackage{graphicx}
\pdfoutput=1 
\usepackage{mathrsfs}
\usepackage[T1]{fontenc} 
\usepackage{bbold}
\usepackage{url}
\usepackage{hyperref}
\usepackage[english]{babel}
\usepackage[utf8]{inputenc}
\usepackage{amsmath}
\usepackage{color}
\usepackage{amsfonts}
\usepackage{amssymb}
\usepackage{mathtools}
\usepackage{placeins}
\usepackage{float}
\usepackage{tabularx}
\usepackage{adjustbox}
\usepackage{caption}
\usepackage{subcaption}

\title{\boldmath Non-perturbative stabilization of two K\"ahler moduli in type-IIB/F theory and the inflaton potential}
\shorttitle{\boldmath Non-perturbative stabilization of two K\"ahler moduli in type-IIB/F theory and the inflaton potential} 

\author{Abhijit Let\inst{} \and Arunoday Sarkar\inst{} \and Chitrak Sarkar\inst{} \and Buddhadeb Ghosh \inst{}}
\institute{                    
  \inst{} Centre of Advanced Studies, Department of Physics, The University of Burdwan,\\Burdwan 713 104, India
  
}

\pacs{11.25.Wx}{String and brane phenomenology}
\pacs{98.80.Qc}{Quantum cosmology}
\pacs{98.80.Cq}{Particle-theory and field-theory models of the early Universe (including cosmic pancakes, cosmic strings, chaotic phenomena, inflationary universe, etc.)}

\abstract{ We consider a combination of perturbative and non-perturbative corrections in K\"ahler moduli stabilizations in the configuration of three magnetised intersecting D7 branes in the type-IIB/F theory, compactified on the 6d $T^6/Z_N$ orbifold of Calabi-Yau three-fold ($CY_3$). Two of the K\"ahler moduli are stabilized non-perturbatively, out of the three which get perturbative corrections up to one-loop-order multi-graviton scattering amplitudes in the large volume scenario. In this framework, the $dS$ vacua are achieved through all K\"ahler moduli stabilizations by considering the $D$-term. We obtain inflaton potentials of slow-roll plateau-type, which are expected by recent cosmological observations. Calculations of cosmological parameters with the potentials yield experimentally favoured values.}

\begin{document}

\maketitle
\flushbottom
The prime motivation of moduli stabilization in string theory \cite{Becker:2006dvp} is to provide a four-dimensional UV-finite effective theory from a ten-dimensional string compactification scenario. Such a theory should resolve the inadequacies of the standard model of particle physics and concomitantly serve as a consistent foundation for the cosmological inflation in the early universe. The latter is more robust not only for explaining the observational CMB anisotropies, polarizations and large-scale structure surveys but also for verifying the string theory and its ingredients from the cosmological point of view. The current frontier of theoretical physics is looking for the major developments in an inflationary model building in the arena of string cosmology.\par  In particular, it is imperative to know whether an inflaton potential is derivable in the framework of string theory through K\"{a}hler moduli stabilizations using various quantum corrections of perturbative and non-perturbative origins \cite{Antoniadis:2018ngr,Antoniadis:2020stf,Conlon:2005jm,Let:2022fmu}. This is a challenging task because the possibility of obtaining a (meta)stable $dS$ vacuum as an effective description of cosmology in four dimensions is still debatable in so far as the reliability of the quantum corrections vis-\`a-vis the various proposed conjectures such as the swampland (see, for example, \cite{Agrawal:2018own,Palti:2019pca,Junghans:2018gdb,Akrami:2018ylq,Denef:2018etk,Conlon:2018eyr}) is considered. Nonetheless, very recently this issue has been rejuvenated in Refs.\cite{Basiouris:2020jgp,Basiouris:2021sdf,Let:2022fmu} by showing that the combined effects of the perturbative \cite{Antoniadis:2019doc,Antoniadis:2018ngr,Antoniadis:2018hqy,Antoniadis:2020stf,Antoniadis:2019rkh} and the non-perturbative \cite{Kachru:2003aw,Haack:2006cy,Bianchi:2011qh,Bena:2019mte} quantum corrections can be a viable path towards realizing a cosmologically favourable stable $dS_4$ vacuum in the string theory. Specifically, Ref. \cite{Basiouris:2021sdf} illustrates that several coupling regimes among the non-perturbatively stabilized K\"{a}hler moduli exist, which can be adapted to redefine the parameters of the final effective potential. But, very few efforts have been made to achieve the slow-roll plateau and to correctly understand the corresponding cumulative quantum dynamics of the K\"{a}hler moduli fields with these couplings.\par 
 In our recent work \cite{Let:2022fmu}, we have shown that if we stabilize one of the three K\"{a}hler moduli fields non-perturbatively corresponding to three intersecting $D7$ branes in type-IIB/F theory compactified on $T^6/\mathbb{Z}_N$ orbifold and other two only perturbatively, then after a suitable canonical normalization procedure, we can get a slow-roll inflaton potential in the $dS$ space. The efficacy of this potential has been checked \cite{Let:2022fmu} by comparing the calculated constraints of inflation with those of the Planck-2018 data \cite{Planck:2018jri,Planck:2018vyg}. We also mention here that there is an auxiliary field in this formalism, which is stabilized at the bottom of the potential.
 \par 
  We extend, here, our study by stabilizing two of the K\"{a}hler moduli non-perturbatively. Then, we examine the possible parameter spaces for the origin of the slow-roll inflaton potential by analysing various moduli couplings and the various manifolds. The implications of this generalization will be discussed.\par
  We have considered, here, a geometrical configuration which is based on the model proposed in Refs. \cite{Antoniadis:2018hqy,Basiouris:2020jgp,Antoniadis:2018ngr}, containing three magnetized  $D7$ branes wrapping intersecting four-cycles. The complexified volume modulus, contains the K\"ahler moduli $\tau_k$, whose simplified form is given in \cite{Giddings:2001yu}, as,
   \begin{equation}
       \rho_k = b_k +i \tau_k, \quad k=1,2,3, 
       \label{rho}
   \end{equation} where,
   $b_k$ is a four-cycle axion coming from the $\boldmath RR$ sector in type-IIB theory and $\tau_k$ is a K\"ahler modulus which corresponds to the volume of the four-cycle wrapped by the $D7$ brane. The imaginary part of the volume modulus gives the volume of the $CY_3$ and it is given in \cite{Let:2022fmu,Basiouris:2021sdf} as,
 \begin{equation}
     \mathcal{V} = \sqrt{\tau_1 \tau_2 \tau_3} =\sqrt{\frac{i}{8}(\rho_1 -\Bar{\rho}_1)(\rho_2-\Bar{\rho}_2)(\rho_3-\Bar{\rho}_3)} .
     \label{Eq:2.1}
 \end{equation}
  Type-IIB string theory contains three-form fluxes $ G_3 = F_3 - S H_3$, where $S=C_0 + ie^{-\phi}$ is known as axion-dilaton (AD) moduli field, $\phi$ is called dilaton field which is related to string coupling constant as $g_s = e^\phi$ and $F_3 =dC_2, $\quad$ $$ H_3 =dB_2$ are the three-form field strengths. Here $C_0, C_2$ are zero- and two-  form potentials respectively and $B_2$ is a Kalb-Ramond field. The fluxes around the compactified cycles in $\boldmath CY_3$ generate the flux-induced superpotential which is given in \cite{Gukov:1999ya}, as,
 \begin{equation}\small
    \mathcal{W}_0 (S,z_\alpha)  = \int_{\chi_{_6}} G_3 (S,z_\alpha) \wedge \Omega {(z_\alpha) }, 
    \label{Eq:2.4}
 \end{equation}
  where  $\Omega {(z_\alpha)}$ represents the (3,0) holomorphic form which is a function of complex-structure (CS) moduli fields ($z_\alpha)$, `$\alpha$' runs over all CS moduli ($\alpha = 1,2,........h^{2,1}; \quad   h^{2,1}$ is a non-vanishing hodge number of $CY_3$). The tree-level superpotential (\ref{Eq:2.4}) is independent of K\"ahler moduli. \par 
 There is another important ingredient in string theory known as K\"ahler potential which generates the metric of moduli (CS and K\"ahler structure (KS)) spaces of $ CY_3$,  $\it{i.e.}$, $K_{I\Bar{J}} =\partial_I \partial_{\Bar{J}}\mathcal{K}_0$, where `I', `J' run over all moduli fields (CS, KS and AD), and $\mathcal{K}_0$ is tree level K\"ahler potential which depends logarithmically on all moduli fields \cite{Basiouris:2021sdf,Let:2022fmu}, as,
 \begin{equation}\small
 \begin{split}
     \mathcal{K}_0 =-2 \ln \Big(\sqrt{\frac{(\rho_1 -\Bar{\rho}_1)(\rho_2-\Bar{\rho}_2)(\rho_3-\Bar{\rho}_3)}{ \left(2i\right)^3}}\Big)\\ - \ln \big(-i(S-\Bar{S}\big) - \ln \Big(-i\int_{\chi_{_6}}\Omega\wedge\Bar{\Omega}\Big). 
     \label{eq:2.6}
     \end{split}
 \end{equation}\par
 In order to compute the F-term potential for various moduli fields in $D=4,\  \mathcal{N}=1$ supergravity theory, a standard formula has been given in \cite{Giddings:2001yu}:
 \begin{align}\small
     V_F =e^{\mathcal{K}} \sum_{I,J} \Big(K^{I\Bar{J}}\mathcal{D}_I \mathcal{W}\mathcal{D}_{\Bar{J}}\mathcal{\Bar{W}}-3| \mathcal{W}|^2 \Big),
     \label{Eq:2.7}
 \end{align}
 where $K^{I\Bar{J}}$ is the inverse metric of $K_{I\bar{J}}$ and $\mathcal{D}\mathcal{W}=\partial \mathcal{W}+\left({\partial\mathcal{K}}\right)\mathcal{W}$ is a covariant derivative. At tree-level, the F-term potential for K\"ahler moduli vanishes and we obtain the F-term potential from eq. (\ref{Eq:2.7}) for all CS and the dilaton moduli fields, i.e.,
 \begin{equation}\small
     V_{F}=e^{\mathcal{K}_0} \sum_{a,b} K^{a,\Bar{b}}\mathcal{D}_a \mathcal{W}_0\mathcal{D}_{\Bar{b}}\mathcal{\Bar{W}}_0,
     \label{Eq:2.8}
 \end{equation}
 where, a,b label the indices of moduli fields, excluding the K\"ahler ones. Using the supersymmetric conditions ($\mathcal{D}_S \mathcal{W}_0=0$ and  $\mathcal{D}_{z_\alpha} \mathcal{W}_0 = 0$), all CS and dilaton moduli are fixed with large masses        \cite{Kachru:2003aw,Giddings:2001yu}, but K\"ahler moduli remain unstabilized. We now begin discussing the stabilization of the K\"ahler moduli.  \par
 At first, we focus on the non-perturbative modification of $\mathcal{W}_0$. In this context, two mechanisms viz., Euclidian D3 branes instantons and gaugino condensations from stacks of D7 branes have been proposed by KKLT \cite{Kachru:2003aw}. Taking either effect, the superpotential becomes K\"ahler moduli dependent, as,
 \begin{align}\small
     \mathcal{W} = \mathcal{W}_0 +\sum_k A_k e^{i a_k\rho_k}, 
 \label{sup-1}
 \end{align}
 where `$k$' depends on the number of smaller K\"ahler moduli as against other moduli which are suppressed exponentially, in a model-dependent way. $\mathcal{W}_0$ is flux-induced tree-level superpotential defined in (\ref{Eq:2.4}) which is considered to be a real constant. The coefficient $A_k$, also a real constant \cite{Kachru:2003aw}, is a function of CS moduli and dilaton \cite{Cicoli:2007xp} (In the case of gaugino condensation on stack of D7 branes, wrapping four-cycles, $A_k$ depends on CS and dilaton and open-string moduli associated with the D7 branes\cite{Haack:2006cy}). $a_k$ is a real constant, whose numerical values are $2\pi$ for instanton effect and ($2\pi/N_k$) for gaugino condensation, where $N_k$ is the rank of the gauge group linked with the branes. In general, the non-perturbative effects act on the K\"ahler moduli in all the directions of the moduli space.  For example, in the KKLT model, all  K\"ahler moduli appear in the superpotential \cite{AbdusSalam:2020ywo,Moritz:2017xto}. \par 
 In our earlier work\cite{Let:2022fmu}, non-perturbative contribution from a single modulus in the superpotential has ensured stable $dS$ vacua and a plateau-type potential, which describes the slow-roll inflation. However, due to the variety of compactification manifolds, a more likely scenario is that more than one K\"ahler moduli contribute to the non-perturbative correction in the superpotential. The present paper aims to take two non-perturbative terms in the superpotential, which come from K\"ahler moduli, $\rho_1$ and $\rho_2$ \cite{Basiouris:2021sdf} and examine the consequences of this procedure on the overall behaviour of the inflaton potential. The non-perturbatively corrected superpotential in eq.(\ref{sup-1}) becomes,
 \begin{equation}\small
     \mathcal{W}= \mathcal{W}_0 + A_1 e^{ia_1\rho_1} +A_2 e^{ia_2\rho_2} . 
\label{non2}
\end{equation}
The remaining modulus ($\rho_3$) is assumed to be large, so that we can work within a large volume scenario (LVS)\cite{Balasubramanian:2005zx}. It is described in \cite{Burgess:2005jx} that the flux-induced superpotential ($\mathcal{W}_0$) in string theory can not receive perturbative corrections due to a non-renormalization theorem. The volume term of K\"ahler potential ($\mathcal{K}_0$) in eq. (\ref{eq:2.6}) can receive various types of perturbative corrections, $\it{e.g.}$, $\alpha'$- and logarithmic one-loop corrections. In \cite{Becker:2002nn}, the authors described the $\alpha'^3$ contribution to the K\"ahler potential and computed the modified k\"ahler potential in type-IIB theory by using mirror symmetry, which changes the volume with a constant shifting term $\boldmath\xi =- \zeta(3)\chi/4(2\pi)^3g_s^{3/2}$, where $\chi$ is the Euler number of $\boldmath CY_3$. The one-loop multi-graviton scattering amplitude in type-IIB string theory gives rise to the second type of quantum correction \cite{Antoniadis:2019rkh} to the K\"ahler potential. However, the low-energy expansions \cite{Green:1999pv, Green:2010wi} of multi-graviton scattering amplitude generate the $\boldmath R^4$ coupling terms \cite{Antoniadis:1997eg,Antoniadis:2019rkh}, whose coefficients contain both the tree-level as well as the one-loop contributions. Considering only the gravitational  sector, the supergravity action with quantum corrections upto one-loop order in type-IIB string theory is given by \cite{Antoniadis:2002tr,Antoniadis:2019rkh,Basiouris:2021sdf},
 \begin{equation}
 \begin{split}\small
     S_{grav} =\frac{1}{{(2\pi)}^7 \alpha'^4}\int_{\mathcal{M}_4 \times {\chi}_{_6}} e^{-2\phi}\mathcal{R}_{(10)}+ \frac{\chi}{{(2\pi)}^4 \alpha'} \\ \times\int_{\mathcal{M}_{4}}\Big(2\zeta(3)e^{-2\phi} +4\zeta(2) \Big) {\mathcal{R}_{(4)}},
 \label{Eq:2.11}
 \end{split}
 \end{equation}
 where, the terms involving the Riemann zeta functions $\zeta(3), \zeta(2)$ come respectively from the tree-level and the one-loop scattering amplitudes, after low energy expansions \cite{Green:1999pv}. $\mathcal{R}_{(d)}$ is defined as the d-dimensional Ricci scalar.  The localized EH term $(\mathcal{R}_{(4)}=\boldmath R \wedge \boldmath e^2)$ \cite{Antoniadis:2002tr,Antoniadis:2019rkh}, appears in four dimensions through dimensional reduction, for non-vanishing Euler number of $\boldmath CY_3$. Here, $\it{e}$ is a `vielbein'. The Euler number in eq. (\ref{Eq:2.11}) is given by,
 \begin{equation}\small
     \chi=\frac{3!}{(2\pi)^3} \int_{\chi_{_6}} R \wedge  R \wedge R ,
 \end{equation}
 where, $\boldmath R$ is the 2-form Ricci curvature. Orbifold fixed points in type-IIB string theory, compactified on $T^6/\mathbb{Z_N}$, which are associated with vertices for non-zero Euler numbers, are the sources of localized gravity. The localized sources emit massless gravitons and massive $KK$ excitations, which propagate along all directions in the compactified $6d$ space. The width ($\omega$) of the localized gravity implicitly depends on Euler number of $\boldmath CY_3$ as \cite{Antoniadis:2019rkh},  
 \begin{equation}\small
     \omega\approx \frac{l_s}{\sqrt{N}} ,
 \label{Eq:width}
 \end{equation}
 where  $l_s$ is the fundamental string length and the integer N is related to the Euler number of $\boldmath CY_3$ ( $|\chi|\sim N$). The $KK$ modes are exchanged between localized gravity sources and the D7-branes/O7 planes, which can give rise to one-loop logarithmic correction   \cite{Antoniadis:2019rkh}, and the supergravity action eq. (\ref{Eq:2.11}) takes the form
 \begin{equation}\small
 \begin{split}
     S_{grav} =\frac{1}{{(2\pi)}^7 \alpha'^4}\int_{\mathcal{M}_4 \times {\chi}_{_6}} e^{-2\phi}\mathcal{R}_{(10)} \\+ \frac{\chi}{{(2\pi)}^4 \alpha'}\int_{\mathcal{M}_{4}}(2\zeta(3)e^{-2\phi}\\+4\zeta(2)(1 -\sum_k e^{2\phi}T_k \log(\frac{R^k_\perp}{\omega}))) {\mathcal{R}_{(4)}},
 \label{Eq:mods}
 \end{split}
 \end{equation}
 where, $T_k$ is the tension of the k-th D7 brane, and $R_\perp$ stands for the size of the two-dimensional space transverse to the D7 branes. For simplicity, tensions of all D7 branes are assumed to be the same, ($\it{i.e.}$, $T_k$=T=$e^{-\phi}T_0$). The quantum corrected term ($\delta$) can be deduced from eq. (\ref{Eq:mods}), which is written as\cite{Antoniadis:2019rkh,Basiouris:2021sdf,Antoniadis:2018hqy},
 \begin{equation}\small
\delta=\xi+\sum_{j=1}^3\eta_j\ln{(\tau_j)}=\xi+\eta\ln{(\tau_1\tau_2\tau_3)},
 \label{Eq:delta}
 \end{equation}
 where, the coefficient $\eta_j\equiv\eta=-\frac{1}{2}g_sT_0\xi$ is a constant and $\xi$ is associated with $\chi$: $\xi=-\frac{\zeta(2)g_s^2}{2}\chi$ for orbifolds and $\xi=-\frac{\zeta(3)}{4}\chi$ for the smooth $CY_3$ ($\it{i.e.}$, the non-compact limit). The term $\delta$ in eq. (\ref{Eq:delta}) modifies the tree level K\"ahler potential and the volume term become \cite{Basiouris:2021sdf, Antoniadis:2019rkh},
 \begin{equation}\small
     \mathcal{K}(\rho_1,\rho_2,\rho_3)=-2\ln{(\mathcal{V}+\xi+2\eta \ln{\mathcal{V}})}=-2\ln{\mathcal{Y}},
 \label{k2}
 \end{equation} where, $\mathcal{Y}=(\mathcal{V}+\xi+2\eta \ln{\mathcal{V}})$.
 The no-scale structure ($\it{i.e.}$, $\sum_{k,k'=1}^3 {\mathcal{K}_0}^{{\rho_k}\bar{\rho}_{k'}} \partial_{\rho_k} \mathcal{K}_0 \partial_{\bar{\rho}_{k'}} \mathcal{K}_0 = 3$) is broken by the K\"ahler potential eq. (\ref{k2}), as,
 \begin{equation}\small
    \sum_{k,k'=1}^3 {\mathcal{K}}^{{\rho_k}\bar{\rho}_{k'}} \partial_{\rho_k} \mathcal{K} \partial_{\bar{\rho}_{k'}} \mathcal{K} \neq 3.
    \label{no-scale}
\end{equation}
Eq.(\ref{no-scale})  implies a non-vanishing dependence of $F$-term potential on K\"ahler moduli and the eq. (\ref{Eq:2.7}) becomes, 
\begin{equation}\small
    V_F = e^{\mathcal{K}}\left(\sum_{k,k'=1}^{3} {\mathcal{K}}^{\rho_k \bar{\rho}_{k'}} \mathcal{D}_{\rho_k} \mathcal{W} \mathcal{D}_{\bar{\rho}_{k'}}\bar{\mathcal{W}} - 3\mathcal{W} \bar{\mathcal{W}}\right),
    \label{F}
\end{equation}
using eqs. (\ref{F}),(\ref{non2}) and (\ref{k2}), the F-term potential has been computed in WOLFRAM MATHEMATICA 12, and the exact forms are obtained as,
\begin{equation}\small
\begin{split}
     V_F^{(p)}&=e^\mathcal{K}\sum_{j,\Bar{{j'}}}\left(\mathcal{K}^{\rho_j{\Bar{\rho_{j'}}}}\partial_{\rho_j}\mathcal{K}\partial_{\Bar{\rho_{j'}}}\mathcal{K}-3\right)|\mathcal{W}_0|^2\\
     &=\frac{-3\mathcal{W}_0^2\left(8\mathcal{V}\eta+4\eta^2+4\eta\xi+8\eta^2\ln\mathcal{V}-\mathcal{V}\xi-2\mathcal{V}\eta\ln\mathcal{V}\right)}{\mathcal{Y}^2\left(2\mathcal{V}^2+16\eta\mathcal{V}+12\eta^2+\left(4\eta-\mathcal{V}\right)\left(\xi+2\eta\ln{\mathcal{V}}\right)\right)}
     \label{Fp}
\end{split}
\end{equation}
and
\begin{multline}\small
    V_F^{(r)}=\frac{8\left(a_1^2\tau_1^2\Tilde{A_1}^2+a_2^2\tau_2^2\Tilde{A_2}^2\right)}{\mathcal{Y}\left(\mathcal{V}+2\eta\right)} \\ \times\left(\frac{\left(\mathcal{V}^2+6\mathcal{V}\eta+4\eta^2+2\eta\left(\xi+2\eta\ln\mathcal{V}\right)\right)}{\left(2\mathcal{V}^2+16\eta\mathcal{V}+12\eta^2+\left(4\eta-\mathcal{V}\right)\left(\xi+2\eta\ln{\mathcal{V}}\right)\right)}\right) \\
    + \left(\frac{8\mathcal{W}\left(a_1\tau_1\Tilde{A_1}+a_2\tau_2\Tilde{A_2}\right)}{\mathcal{Y}^2\left(2\mathcal{V}^2+16\eta\mathcal{V}+12\eta^2+\left(4\eta-\mathcal{V}\right)\left(\xi+2\eta\ln{\mathcal{V}}\right)\right)}\right)\\ \times\left(\mathcal{V}^2+6\mathcal{V}\eta+4\eta^2+2\eta\left(\xi+2\eta\ln\mathcal{V}\right)\right) + 8\mathcal{W}\\ \times\left(\frac{\left(a_1\tau_1\Tilde{A_1}+a_2\tau_2\Tilde{A_2}\right)\left(-4\mathcal{V}\eta-4\eta^2+\mathcal{V}\xi+2\mathcal{V}\eta\ln\mathcal{V}\right)}{\mathcal{Y}^2\left(2\mathcal{V}^2+16\eta\mathcal{V}+12\eta^2+\left(4\eta-\mathcal{V}\right)\left(\xi+2\eta\ln{\mathcal{V}}\right)\right)}\right)\\+\frac{8\mathcal{W}a_1a_2\tau_1\tau_2\Tilde{A_1}\Tilde{A_2}}{\mathcal{Y}\left(\mathcal{V}+2\eta\right)}\\ \times\left(\frac{\left(-4\mathcal{V}\eta-4\eta^2+\mathcal{V}\xi+2\mathcal{V}\eta\ln\mathcal{V}\right)}{\left(2\mathcal{V}^2+16\eta\mathcal{V}+12\eta^2+\left(4\eta-\mathcal{V}\right)\left(\xi+2\eta\ln{\mathcal{V}}\right)\right)}\right)\\ +\left(\frac{-3\left(8\mathcal{V}+4\eta+4\eta\xi+8\eta^2\ln\mathcal{V}-\mathcal{V}\xi-2\mathcal{V}\ln\mathcal{V}\right)}{\mathcal{Y}^2\left(2\mathcal{V}^2+16\eta\mathcal{V}+12\eta^2+\left(4\eta-\mathcal{V}\right)\left(\xi+2\eta\ln{\mathcal{V}}\right)\right)}\right) \\ \times\left(\mathcal{W}^2-\mathcal{W}_0^2\right).
    \label{Fr}
\end{multline}
In the derivations of eqs.(\ref{Fp}) and (\ref{Fr}), we have assumed $b_k$'s (see eq.(\ref{rho})) to be zero. The indices `p' and `r' respectively refer to purely perturbative and remaining terms and $\Tilde{A_1}=A_1e^{-a_1\tau_1}$, $\Tilde{A_2}=A_2e^{-a_2\tau_2}$. 
At the large volume approximations ($\it{i.e}$,$\mathcal{V}\gg \eta,\xi$), expanding the eqs. (\ref{Fp}) and (\ref{Fr}) and neglecting the terms $\mathcal{O}(\xi^2)$, $\mathcal{O}(\eta^2)$, and $\mathcal{O}(\frac{1}{\mathcal{V}^5})$, we get the separate results for  non-perturbative, perturbative and mixing terms, as,
\begin{multline}\small
     V_F^{(np)}=\frac{4}{\mathcal{V}^2} \Big(a_1\tau_1\Tilde{A_1} \big(\big(1+a_1\tau_1 \big)\Tilde{A_1}+\big(\mathcal{W}_0+\Tilde{A_2}\big)\big)\\+a_2\tau_2\Tilde{A_2} \big(\big(1+a_2\tau_2\big)\Tilde{A_2}+\big(\mathcal{W}_0+\Tilde{A_1}\big)\big)\Big),
\end{multline}

\begin{equation}\small
    V_F^{(p)}=\frac{3\mathcal{W}_0}{2\mathcal{V}^3}\left(\xi-2\eta\left(4-\ln\mathcal{V}\right)\right)-\frac{9\mathcal{W}_0}{\mathcal{V}^4}\eta\xi\ln\mathcal{V},
\end{equation}
\begin{multline}\small
    V_F^{(m)}=\Tilde{A_1}\left(\mathcal{F}_1\Tilde{A_1}+\mathcal{G}_1\mathcal{W}_0\right)  +\Tilde{A_2}\left(\mathcal{F}_2\Tilde{A_2}+\mathcal{G}_2\mathcal{W}_0\right)\\ + \mathcal{H}\Tilde{A_1}\Tilde{A_2},
\end{multline}
where,
\begin{multline}\small
\mathcal{F}_1=\frac{1}{2\mathcal{V}^3}\big(\left(\xi+2\eta\ln\mathcal{V}\right)\left(3+2a_1\tau_1\right)\left(1-2a_1\tau-1\right)\\-8\eta\left(4a_1^2\tau_1^2+6a_1\tau_1+3\right)\big)\\ + \frac{\eta\xi\left(3+a_1\tau_1\right)\left(\left(3-2a_1\tau_1\right)\ln\mathcal{V}+2a_1\tau_1\right)}{\mathcal{V}^4},
\end{multline}
\begin{equation}\small
   \begin{split}
\mathcal{G}_1=\frac{\left(\xi+2\eta\ln\mathcal{V}\right)\left(3-2a_1\tau_1\right)-24\eta\left(1+a_1\tau_1\right)}{\mathcal{V}^3} \\- \frac{6\xi\eta}{\mathcal{V}^4}\big(\left(3-2a_1\tau_1\right)\ln\mathcal{V}+2a_1\tau_1\big),
\end{split} 
\end{equation}
\begin{multline}\small
    \mathcal{F}_2=\frac{1}{2\mathcal{V}^3}\left(\xi+2\eta\ln\mathcal{V}\right)\left(3+2a_2\tau_2\right)\left(1-2a_2\tau-2\right)\\-8\eta\left(4a_2^2\tau_2^2+6a_2\tau_2+3\right)\\ + \frac{\eta\xi\left(3+a_2\tau_2\right)\left(\left(3-2a_2\tau_2\right)\ln\mathcal{V}+2a_2\tau_2\right)}{\mathcal{V}^4},
\end{multline}
\begin{multline}\small
    \mathcal{G}_2=\frac{\left(\xi+2\eta\ln\mathcal{V}\right)\left(3-2a_2\tau_2\right)-24\eta\left(1+a_2\tau_2\right)}{\mathcal{V}^3}\\
    - \frac{6\xi\eta}{\mathcal{V}^4}\big(\left(3-2a_2\tau_2\right)\ln\mathcal{V}+2a_2\tau_2\big),
\end{multline}
\begin{multline}\small
    \mathcal{H}=\frac{1}{\mathcal{V}^3}\Big(\left(\xi+2\eta\ln\mathcal{V}\right)\left(4a_1a_2\tau_1\tau_2-2a_1\tau_1-2a_2\tau_2+3\right) \\
    -8\eta\left(2a_1a_2\tau_1\tau_2+3a_1\tau_1+3a_2\tau_2+3\right)\Big) \\ - \frac{\eta\xi}{\mathcal{V}^4}\Big(\left(\left(8a_1a_2\tau_1\tau_2-12a_1\tau_1-a_2\tau_2-18\right)\ln\mathcal{V}\right) \\
    -4\left(8a_1a_2\tau_1\tau_2+3a_1\tau_1+3a_2\tau_2\right)\Big),
\end{multline}

where the indices `np' and `m' respectively refer to non-perturbative and mixing terms.\par
 The perturbative contributions  $\mathcal{O}(\xi,\eta)$ in eq. (\ref{k2}) can be neglected 
 in large volume limit and the K\"ahler potential in (\ref{k2}) reduces as\cite{Basiouris:2021sdf}, 

\begin{equation}\small
    \mathcal{K}\approx -2\ln\mathcal{V}.
    \label{ka}
\end{equation} The supersymmetric stabilization conditions for the moduli $\rho_1$ and $\rho_2$ are \cite{Basiouris:2021sdf}
\begin{equation}\small
\mathcal{D}_{\rho_1}\mathcal{W}|^{\rho_1 =i\tau_1}_{\rho_2 =i\tau_2}= \mathcal{D}_{\rho_2}\mathcal{W}|^{\rho_1 =i\tau_1}_{\rho_2 =i\tau_2}=0 \ .
\label{eq:super}
\end{equation}

 Now, taking the derivative of (\ref{ka}) with respect to $\rho_1$ and $ \rho_2$, we get,
 \begin{equation}\small
     \partial_{\rho_1}\mathcal{K}=-\frac{1}{\rho_1- \Bar{\rho_1}}; \quad \partial_{\rho_2}\mathcal{K}=-\frac{1}{\rho_2- \Bar{\rho_2}}.
 \label{Eq:kd}
 \end{equation}\par
 Therefore, the covariant derivatives in (\ref{eq:super}) becomes as,
 \begin{equation}\small
 \begin{split}
      \mathcal{D}_{\rho_1}\mathcal{W}|_{\rho_1 =i\tau_1}^{{\rho_2} =i\tau_2}=\frac{\Tilde{A_1}(2a_1\tau_1+1)+\left(\mathcal{W}_0+\Tilde{A_2}\right)}{-2i\tau_1}=0
 \label{Eq:d1}
 \end{split}
 \end{equation}
 \begin{equation}\small
 \begin{split}
     \mathcal{D}_{\rho_2}\mathcal{W}|_{\rho_1 =i\tau_1}^{{\rho_2 =i\tau_2}}=\frac{\Tilde{A_2}(2a_2\tau_2+1)+(\mathcal{W}_0+\Tilde{A_1})}{-2i\tau_2}=0 \ .
 \label{Eq:d2}
 \end{split}
 \end{equation}
 By combining the eqs. (\ref{Eq:d1}) and (\ref{Eq:d2}), we obtain, 
 \begin{equation}\small
     \frac{a_1\tau_1}{a_2\tau_2}e^{-a_1\tau_1}=\frac{A_2}{A_1}e^{-a_2\tau_2}= \beta e^{-a_2\tau_2} , 
 \label{s1}
 \end{equation}
 \begin{equation}\small
     \Big(\frac{a_1\tau_1}{a_2\tau_2}\big(2a_2\tau_2+1\big)+1\Big)e^{-a_1\tau_1}=-\gamma ,
\label{s2}
\end{equation}
 where, $\gamma=\frac{\mathcal{W}_0}{A_1}$ and $\beta=\frac{A_2}{A_1}$ are  constant. The non-perturbative contributions are assumed to be much less than $|\mathcal{W}_0|$\cite{Basiouris:2021sdf} which implies, from eq.(\ref{s1}), a limitation: $\beta e^{-a_2 \tau_2} \ll \gamma$.  In our formalism, we consider two cases:\par
 \underline{First Case}: $a_2\tau_2\gg a_1\tau_1$. Then, from eq. (\ref{s2}), we get
\begin{equation}\small
     w_1=-\left(\frac{1+2a_1\tau_1}{2}\right)=W_{0/-1}\left(\frac{\gamma}{2\sqrt{e}}\right),
     \label{case-1t1}
 \end{equation} similarly, from eq. (\ref{s1}), we get,
 \begin{equation}\small
     -a_2\tau_2=w_2=W_{0/{-1}}\Big(-\frac{a_1\tau_1}{\beta}e^{-a_1\tau_1}\Big) ,
 \label{case-1t2}
 \end{equation}
 where, $W_{0/-1}$ is Lambert $W$-function, with `0' and `-1' corresponding to the upper and the lower branch, respectively.
 The positive values of $a_1\tau_1$ in eq. (\ref{case-1t1}) restricts us to work in the lower branch. Then, from the range of the argument of the  $W_{-1}$ function we get \cite{Basiouris:2021sdf}
 \begin{equation}\small
     -2/\sqrt{e}\leq \gamma < 0 .
 \end{equation}
 Now, in order to get simplified expressions in what follows, we introduce a new parameter:
\begin{equation}\small
    \epsilon=-\frac{1+2 w_1}{w_1} .
    \label{Epsio-1}
\end{equation}
 Using eq. (\ref{Epsio-1}), the F-term potentials together become
 \begin{multline}\small
V_F^{(1)}=\frac{\left(\epsilon\mathcal{W}_0\right)^2}{32w_2^2\mathcal{V}^3}\Big(\left(2w_2+1\right)\left(14w_2+3\right)\left(\xi+2\eta\ln\mathcal{V}\right)\\ -24\eta -16w_2\mathcal{V}\left(1+w_2\right)\\ 
     +2\eta\xi\frac{24w_2-\left(68w_2^2+60w_2+9\right)\ln\mathcal{V}}{\mathcal{V}}\Big), \label{F1}
 \end{multline} where, the superscript `(1)' signifies the first case.
 Here, $w_2=-a_2\tau_2\ll-1$ because $a_1\tau_1\gtrsim \mathcal{O}(1)$, so that higher order non-perturbative contributions are eliminated\cite{Balasubramanian:2004uy}. In this scenario, eq. (\ref{F1}) becomes,
 \begin{equation}\small
   V_F^{(1)}=\left(\epsilon\mathcal{W}_0\right)^2\left(\frac{7\left(\xi+2\eta \ln{\mathcal{V}}\right)-4\mathcal{V}}{8\mathcal{V}^3} - \frac{17\eta\xi\ln{\mathcal{V}}}{4\mathcal{V}^4}\right).
   \label{F1a}
 \end{equation} \par
 \underline{Second case}: $a_1\tau_1\approx a_2\tau_2\gtrsim \mathcal{O}(1)$. Then from eq.\ref{s1} $\beta\approx1$. Now, from eq.(\ref{s2}), we get
 \begin{equation}\small
    w'=-(1+a_1\tau_1)=W_{0/-1}\Big(\frac{\gamma}{2e}\Big).
    \label{case-2 t1,2}
\end{equation}
 Like in the first case we work in the $W_{-1}$ branch, which gives the limiting values as $-2\leq\gamma < 0$. Now, in terms of a new parameter 
 \begin{equation}
     \epsilon'=\frac{1+w'}{w'},
 \end{equation} we get the F-term potential ($\it{cf.}$, eq.\ref{F1}) as,

\begin{equation}\small
V_F^{(2)}=\left(\epsilon'\mathcal{W}_0\right)^2\left(\frac{7\left(\xi+2\eta\ln\mathcal{V}\right)}{2\mathcal{V}^3}-\frac{2}{\mathcal{V}^2}-\frac{17\eta\xi\ln\mathcal{V}}{\mathcal{V}^4}\right),
    \label{F2}
\end{equation} where the superscript `(2)' denotes the second case.\par
\begin{figure}[H]\large
  \centering
\onefigure[width=0.5\linewidth]{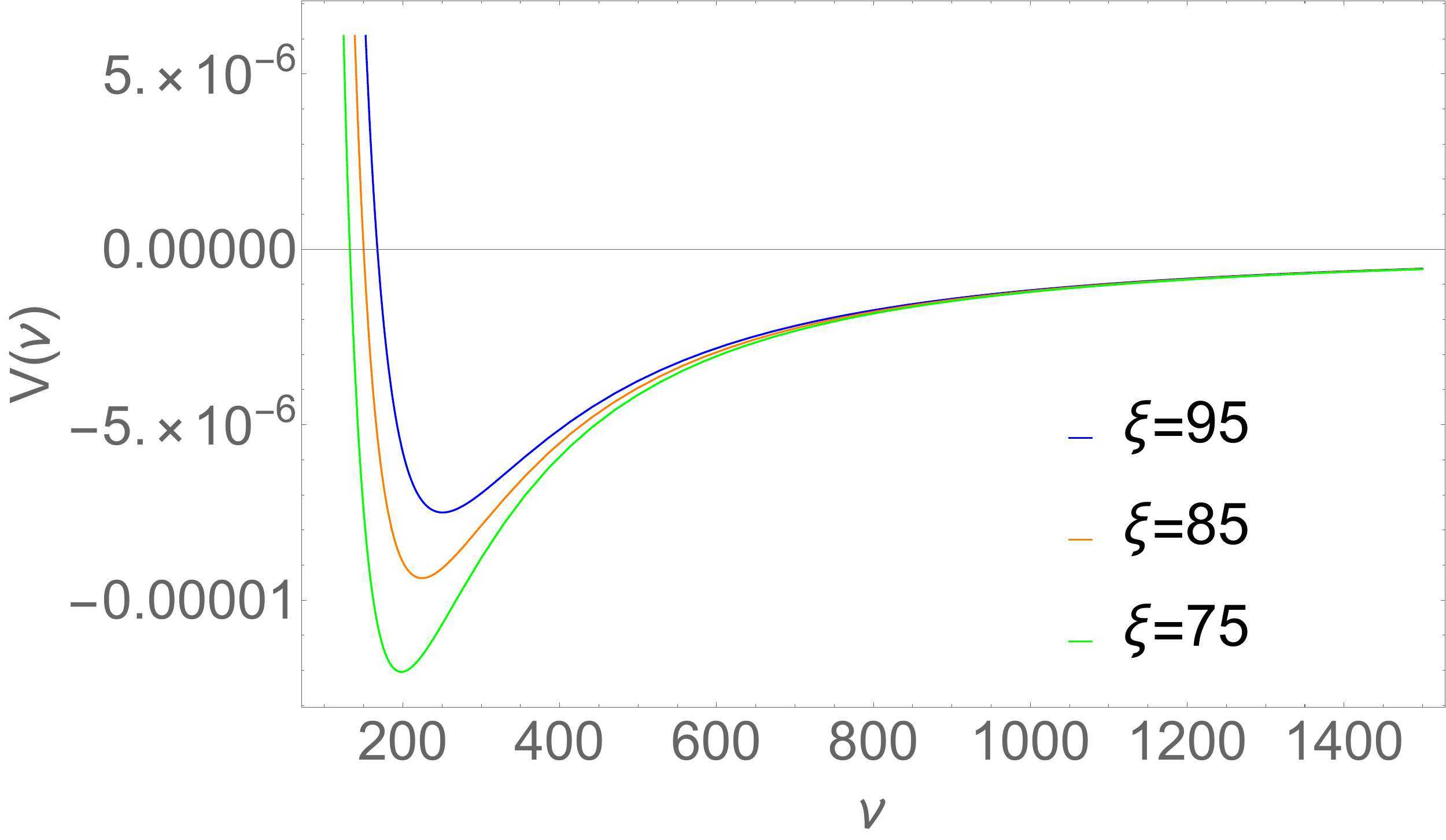}
    \caption{First case: F-term potential eq.(\ref{F1a}) with $AdS$ minima for different values of $\xi$.}
    \label{fig case-1-ads}   
   \end{figure}

\begin{figure}[H]\large
  \centering
    \onefigure[width=0.5\linewidth]{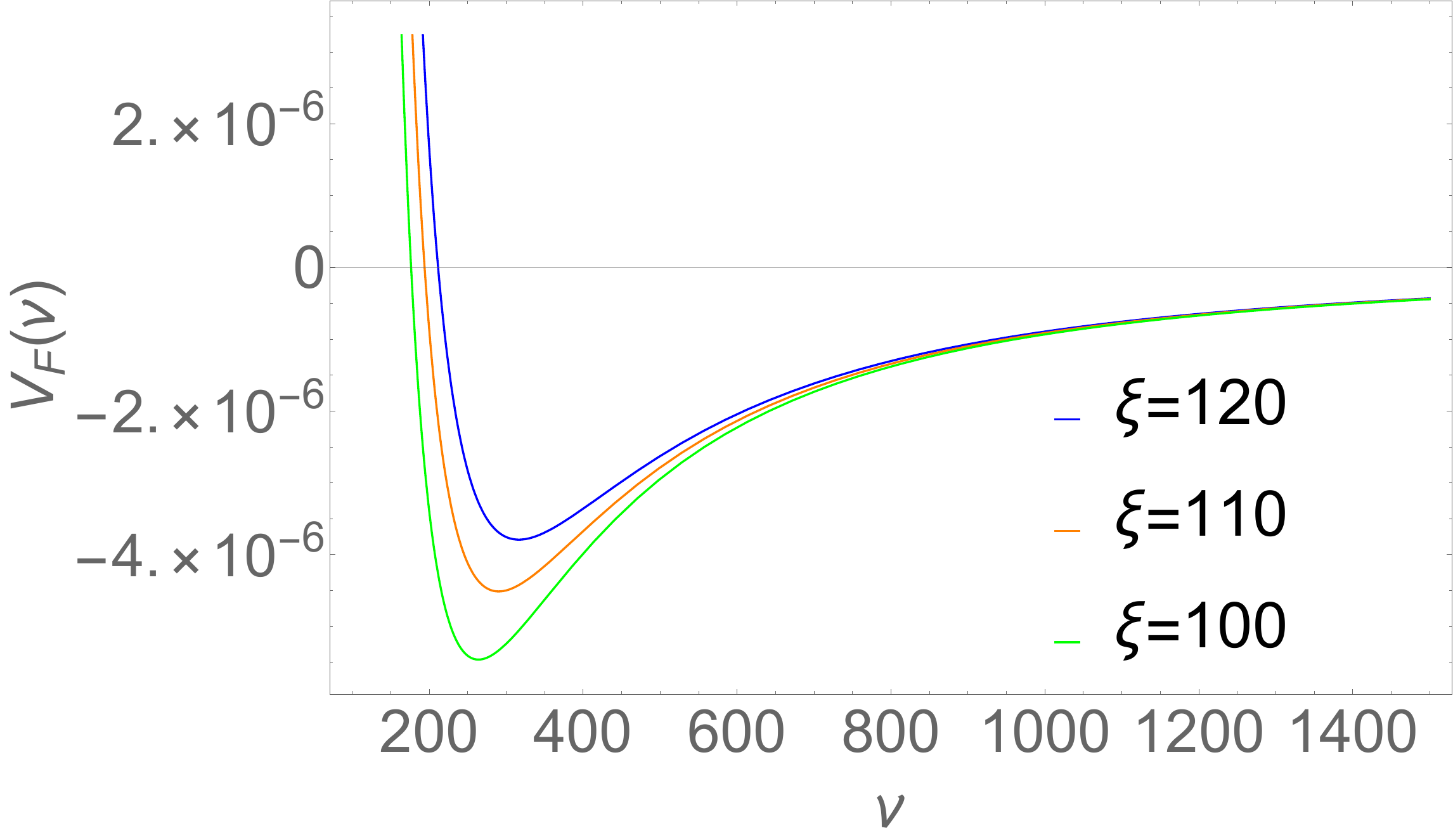}
    \caption{Second case: $F$-term potential eq.(\ref{F2}) with $AdS$ minima for different values of $\xi$}
    \label{fig case-2 ads}   
   \end{figure}

 
The $F$-term potentials in eqs. (\ref{F1a}) and (\ref{F2}) exhibit $AdS$ minima, as shown in Figures (\ref{fig case-1-ads}) and (\ref{fig case-2 ads}) respectively, plotted for three different values of $\xi$. Eqs. (\ref{F1a}) and (\ref{F2}) are not cosmologically very useful because they are in $AdS$ space. \par
  To get $dS$ vacua, an uplifting agent \cite{Kachru:2003aw,Balasubramanian:2005zx,AbdusSalam:2020ywo} have to be added in $F$-term. In the present geometrical set-up, the world volume fluxes along the $D7$ branes, when the fluxes are associated with $U(1)$ gauge fields, naturally, give rise to the $D$-term contribution \cite{Burgess:2003ic,Basiouris:2021sdf,Antoniadis:2018hqy} which uplifts the $AdS$ to a $dS$ vacuum. The original formula of $D$-term potential in the intersecting $D7$ branes is given in \cite{Antoniadis:2018hqy,Cremades:2007ig,Haack:2006cy,Burgess:2003ic} and we use the $D$-term potential, which takes the form \cite{Antoniadis:2018hqy,Basiouris:2020jgp}, as,
\begin{equation}\small
\begin{split}
    V_D=\sum_{i=1}^3\frac{g_{i}^2}{2}\left(\sqrt{-1}Q_i\partial_{\rho_i}\mathcal{K}+\sum_{j}q_j|\langle\Phi_j\rangle|^2\right)^2 \approx \sum_{i=1}^3 \frac{d_i}{\tau_i^3},
    \label{D-term}
    \end{split}
\end{equation}
where $g_i$'s ($g_i^{-2}= \tau_i +$ flux and curvature part containing dilaton) stands for gauge coupling of D7 branes, $Q_i$ are the charges of four-cycle volume moduli $\rho_i$ acquired under U(1) as a shift symmetry and $d_i$ are positive constants which are proportional to charges $Q_i^2$  For simplicity, the matter fields $\Phi_j$ are assumed to have zero VEVs\cite{Let:2022fmu, Basiouris:2021sdf}.\par 
The effective potential is the sum of the $F$-term and the $D$-term potentials, $\it{i.e.}$,
\begin{equation}\small
     V_{\mathrm{eff}}=V_F + V_D.
\end{equation}
In the first case, from eqs.(\ref{F1a}),(\ref{D-term}) and (\ref{Eq:2.1}), we get the effective potential as,
\begin{multline}\small
   V_{\mathrm{eff}}^{(1)}=(\epsilon\mathcal{W}_0)^2\left(\frac{7(\xi+2\eta\ln\mathcal{V})-4\mathcal{V}}{8\mathcal{V}^3}-17\eta\xi\frac{\ln\mathcal{V}}{4\mathcal{V}^4}\right)\\+ \frac{d_1}{\tau_1^3}+\frac{d_3}{\tau_3^3}+\frac{d_2\tau_1^3\tau_3^3}{\mathcal{V}^6}.
   \label{V-effb1}
\end{multline}

Similarly. the second case, from eqs. (\ref{F2}), (\ref{D-term}) and (\ref{Eq:2.1}), we get the effective potential as, 
\begin{equation}\small
\begin{split}
    V_{\mathrm{eff}}^{(2)}=(\epsilon'\mathcal{W}_0)^2\left(\frac{7(\xi+2\eta\ln\mathcal{V})-4\mathcal{V}}{2\mathcal{V}^3}-17\eta\xi\frac{\ln\mathcal{V}}{\mathcal{V}^4}\right)\\+\frac{d_1}{\tau_1^3}+\frac{d_3}{\tau_3^3}+\frac{d_2\tau_1^3\tau_3^3}{\mathcal{V}^6}.
    \label{V-effb2}
    \end{split}
\end{equation}In eqs. (\ref{V-effb1}) and (\ref{V-effb2}), we have replaced $\tau_2$, one of the stabilized moduli.    By minimizing eqs. (\ref{V-effb1}) and (\ref{V-effb2}) with respect to $\tau_3$, we get its value at which the potential is minimum:
\begin{equation}\small
    \tau_3^{min}=\left(\frac{d_3}{d_2}\right)^{1/6}\frac{\mathcal{V}}{\tau_1^{1/2}}.
    \label{t3}
\end{equation}Using eq. (\ref{t3}) in eqs.(\ref{V-effb1}) and (\ref{V-effb2}), we get
 \begin{equation}\small
    \begin{split}
V_{\mathrm{eff}}^{(1)}|_{\tau_3^{min}}&=(\epsilon\mathcal{W}_0)^2\left(\frac{7(\xi+2\eta\ln\mathcal{V})-4\mathcal{V}+r}{8\mathcal{V}^3}-\frac{17\eta\xi\ln\mathcal{V}}{4\mathcal{V}^4}\right) \\&+ \frac{d_1}{\tau_1^3},
\end{split}
\label{V-eff1}
\end{equation}
 \begin{equation}\small
     \begin{split}
V_{\mathrm{eff}}^{(2)}|_{\tau_3^{min}}&=(\epsilon'\mathcal{W}_0)^2\left(\frac{7(\xi+2\eta\ln\mathcal{V})-4\mathcal{V}+r'}{2\mathcal{V}^3}-\frac{17\eta\xi\ln\mathcal{V}}{\mathcal{V}^4}\right)\\&+\frac{d_1}{\tau_1^3},
\end{split}
\label{V-eff2}
 \end{equation}
 where, $d=2\sqrt{d_2d_3}$, $r=\frac{8d\tau_1^{3/2}}{(\epsilon\mathcal{W}_0)^2}$ and $r'=\frac{2d\tau_1^{3/2}}{(\epsilon'\mathcal{W}_0)^2}$. In eqs. (\ref{V-eff1}) and (\ref{V-eff2}) all the moduli $\tau_{1,2,3}$ are stabilized and the potentials are functions of $\mathcal{V}$.\par Due to the absence of $\tau_3$ modulus in superpotential, it can not be fixed by the supersymmetric condition. In this way, only one internal coordinate $\it{viz.}$, $\tau_3$ is assumed to be variable along $\mathcal{V}$. Then, we need a suitable transformation of $\tau_3$ modulus to obtain a canonically normalized fields $\phi$ where $\tau_{1,2}$ are considered constant according to eqs. (\ref{Eq:d1}) and (\ref{Eq:d2}), as they are supersymmetrically stabilized. The appropriate transformation is \cite{Antoniadis:2018ngr, Let:2022fmu}, as,
\begin{equation}\small
        \phi=\frac{1}{\sqrt{2}}\ln(\tau_1\tau_2\tau_3)
        =\sqrt{2}\ln\mathcal{V}.
        \label{inf}
\end{equation} The canonical normalized field $\phi$ eq.(\ref{inf}) may act as an inflaton field. The corresponding inflaton potentials in the first case are obtained from eqs. (\ref{V-eff1}) by using eq.(\ref{inf}), i.e., 
\begin{equation}\small
\boxed{V^{(1)}(\phi)=p_1 e^{-\frac{3\phi}{\sqrt{2}}}\left[\phi-q_1 e^{\frac{\phi}{\sqrt{2}}}-s_1\phi e^{-\frac{\phi}{\sqrt{2}}}+x_1\right]+\frac{d_1}{\tau_1^3}}.
\label{V-inf1}
\end{equation} where, $p_1=\frac{7\sqrt{2}\eta\left(\epsilon\mathcal{W}_0
    \right)^2}{8},q_1=\frac{4}{7\sqrt{2}\eta}, s_1=\frac{17\xi}{7},x_1=\frac{r+7\xi}{7\sqrt{2}\eta}$ are constant parameters. In the second case, the inflaton potential is obtained from eqs. (\ref{V-eff2})and (\ref{inf}) as
\begin{equation}\small
\boxed{V^{(2)}(\phi)=p_2 e^{-\frac{3\phi}{\sqrt{2}}}\left[\phi-q_2 e^{\frac{\phi}{\sqrt{2}}}-s_2\phi e^{-\frac{\phi}{\sqrt{2}}}+x_2\right]+\frac{d_1}{\tau_1^3}}.
\label{V-inf2}
\end{equation}
where, $p_2=\frac{7\sqrt{2}\eta\left(\epsilon'\mathcal{W}_0\right)^2}{2},q_2=\frac{4}{7\sqrt{2}\eta},s_2=\frac{17\xi}{7},x_2=\frac{r'+7\xi}{7\sqrt{2}\eta}$. It is shown that the resulting inflaton potentials $V(\phi)$  eqs. (\ref{V-inf1}) and (\ref{V-inf2}) have the same form but different parameters.  The factor $\frac{d_1}{\tau_1^3}$ in the inflaton potentials eqs. (\ref{V-inf1}) and (\ref{V-inf2}) serves as a constant uplifting term and $d_1$ is a positive uplifting constant. In our framework, we choose the parameter space as follows:
\begin{table}[H]
    \centering
    \caption{Parameter sets for the first case. We choose the parameters of variations as in refs.\cite{Let:2022fmu,Basiouris:2020jgp,Basiouris:2021sdf}. We have chosen the values of $\eta$ (-0.2)  and the $d_{1,2,3}$ suitably to uplift the potentials to the $dS$ space.}
    \begin{tabular}{|c|c|c|c|c|c|c|c|}
    \hline
    $\mathcal{W}_0$ & $\epsilon$ & $\xi$ & $d$ & $d_1$ & $a_1$ & $\tau_1$ & $m_\phi$ ($M_P$) \\
    \hline
        -1 & 1.67 & 75 & .974 & .069 & 0.1 & 25 & .0037  \\
        -1 & 1.67 & 85 & .974 & .069 & 0.1 & 25 & .0034  \\
         -1 & 1.67 & 95 & .974 & .069 & 0.1 & 25 & .0032  \\
         \hline
    \end{tabular}
    \label{tab:first case}
\end{table}
\begin{table}[H]
    \centering
    \caption{Parameter sets for the second case.}
    \begin{tabular}{|c|c|c|c|c|c|c|c|}
    \hline
    $\mathcal{W}_0$ & $\epsilon'$ &  $\xi$ & $d$ & $d_1$ & $a_1$ & $\tau_1$ & $m_\phi$ ($M_P$) \\
    \hline
        -1 & 0.75 & 100 & .09 & .128 & 0.1 & 30 & .0040  \\
        -1 & 0.75 & 110 & .09 & .128 & 0.1 & 30 & .0035  \\
         -1 & 0.75 & 120 & .09 & .128 & 0.1 & 30 & .0032  \\
         \hline
    \end{tabular}
    \label{tab:second case}
\end{table}
\begin{figure}[H]\large
  \centering
\onefigure[width=0.5\linewidth]{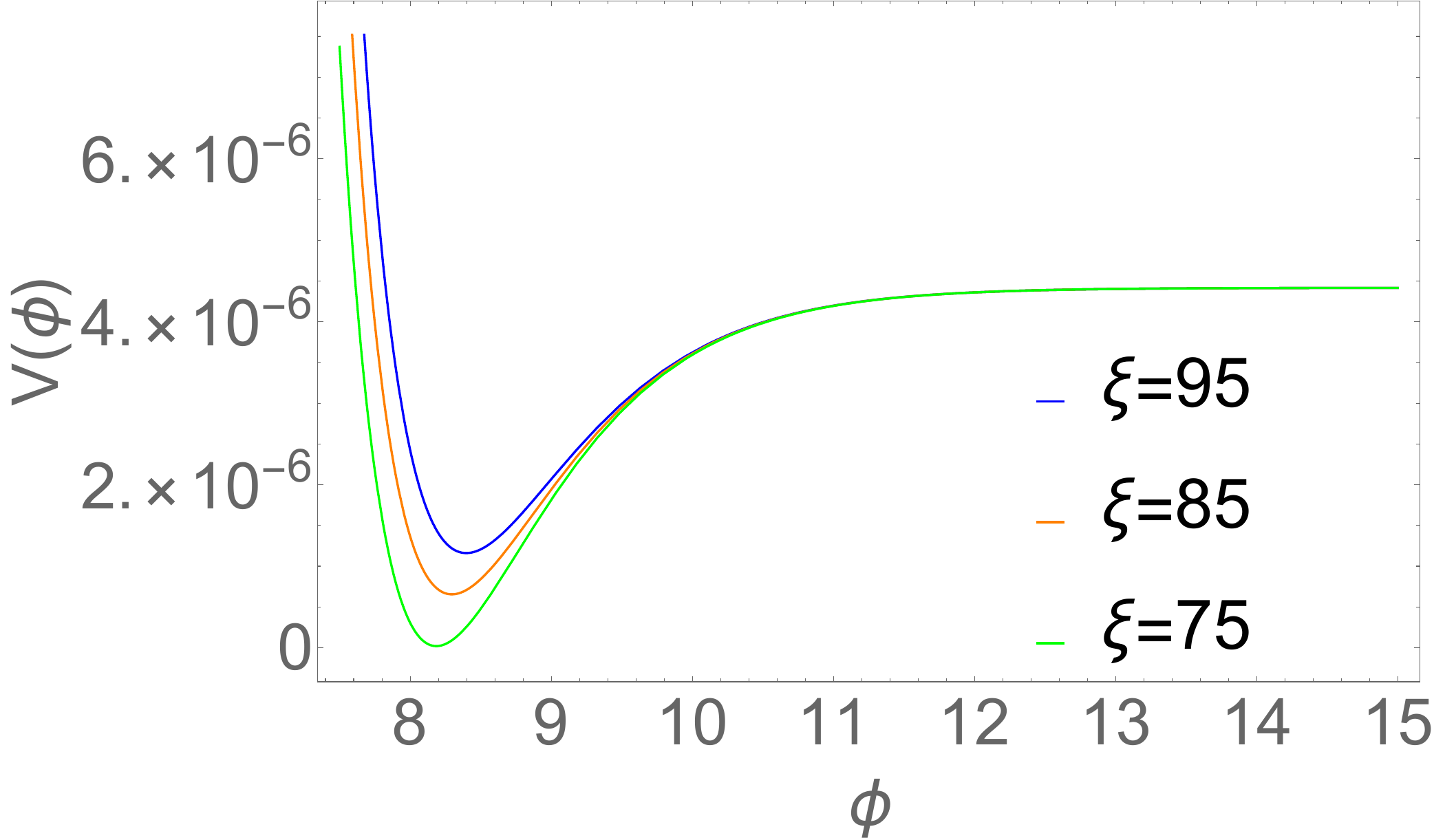}
    \caption{The uplifted inflaton potential eq.(\ref{V-inf1}) with $dS$ minima for different values of $\xi$. The calculated values of $m_\phi$ (in Planck unit) are shown in the last column of Table \ref{tab:first case}}
    \label{fig case-1 inf}   
   \end{figure}
 
\begin{figure}[H]
  \centering
     \onefigure[width=0.5\linewidth]{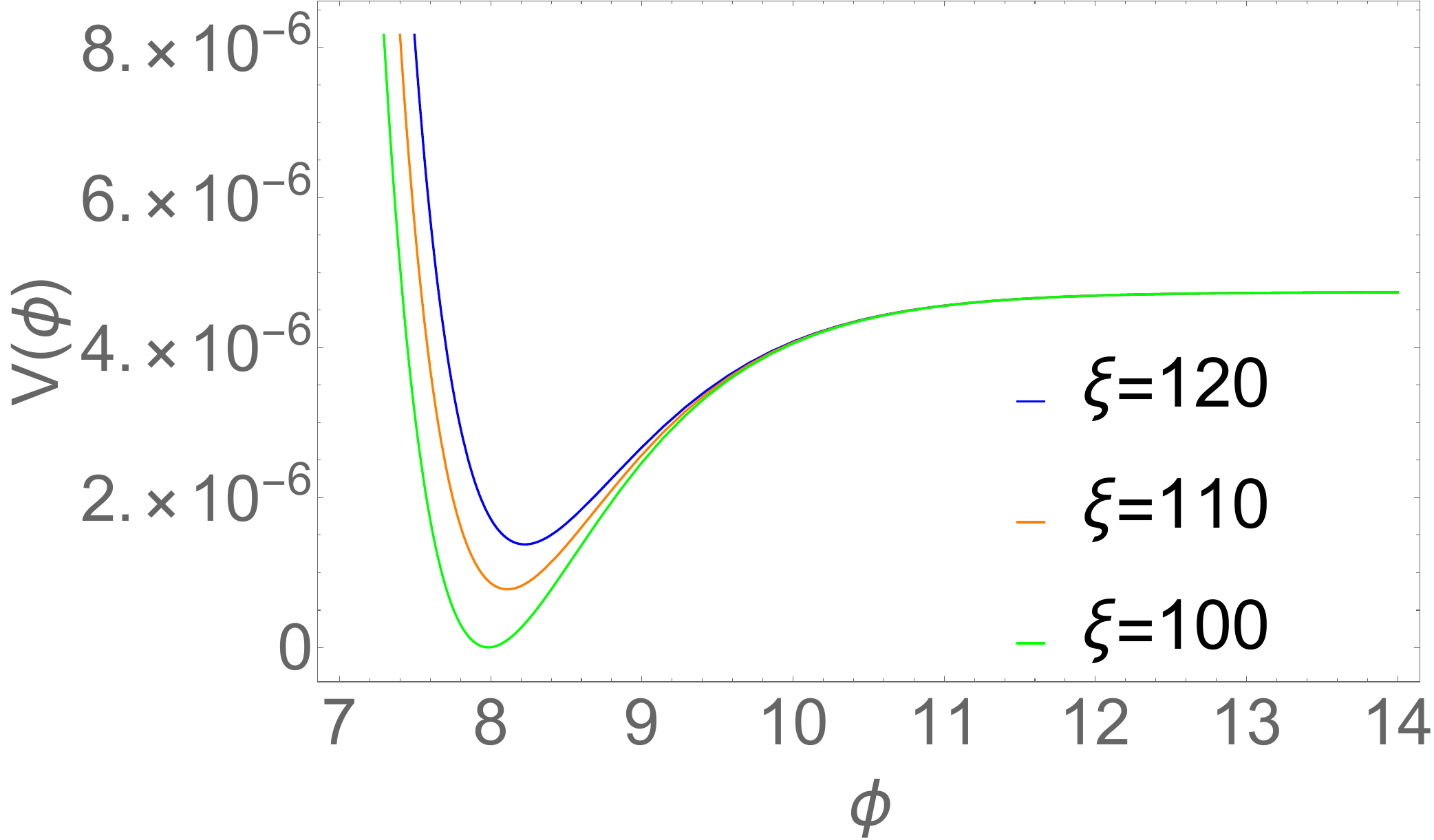}
    \caption{The uplifted inflaton potential eq.(\ref{V-inf2}) with $dS$ minima for different values of $\xi$. The calculated values of $m_\phi$ (in Planck unit) are shown in the last column of Table \ref{tab:second case}}
    \label{fig case-2 inf}   
   \end{figure} 
     
The inflaton potentials eqs. (\ref{V-inf1}) and (\ref{V-inf2}), which are plotted in Figures \ref{fig case-1 inf} and \ref{fig case-2 inf} for three values of $\xi$, describe $dS$ vacua with a plateau nature. Indeed, these plateau nature of inflaton potential arises naturally from the fixing of Kahler moduli ($\tau_{1,2}$) by supersymmetric conditions eqs. (\ref{Eq:d1}) and (\ref{Eq:d2}). \par 
Now, following Ref. \cite{Sarkar:2021ird} we have computed the values of some cosmological parameters for inflaton potentials (eqs. (\ref{V-inf1}) and (\ref{V-inf2})) and presented them in Table 3, along with corresponding experimental data \cite{Planck:2018jri,Planck:2018vyg}.  
\begin{table}[H]
    \centering
    \caption{Cosmological parameters (both calculational and the Planck data\cite{Planck:2018jri,Planck:2018vyg}) at $k=0.002$ Mpc$^{-1}$.}
    \begin{tabular}{|c|c|c|c|}
    \hline
   Parameters  & \vtop{\hbox{\strut POT-1}\hbox{\strut ($\xi=75$)}} & \vtop{\hbox{\strut POT-2}\hbox{\strut ($\xi=100$)}}  & \vtop{\hbox{\strut Planck}\hbox{\strut Value}} \\
          \hline
           $V^{1/4}$ ($M_p$) & $4.58\times 10^{-2}$ & $4.66\times 10^{-2}$ & \vtop{\hbox{\strut $<6.99$}\hbox{\strut $\times 10^{-3}$}}  \\
          \hline
           $n_s$ & 0.9621 & 0.9753 & \vtop{\hbox{\strut $0.9649$}\hbox{\strut $\pm 0.0042$}} \\
        \hline
           $n_t$ & \vtop{\hbox{\strut $-1.81$}\hbox{\strut $\times 10^{-4}$}} & \vtop{\hbox{\strut $-8.09$}\hbox{\strut $\times 10^{-5}$}} & \vtop{\hbox{\strut $<-8$}\hbox{\strut $\times 10^{-3}$}} \\
          \hline
           $r$ & $1.45\times 10^{-3}$ & $6.47\times 10^{-4}$ & <0.064 \\
          \hline
          $N$ & 60.5 & 59.8 & 60-65 \\
          \hline
    \end{tabular}
    \label{tab:second case}
\end{table}

Table 3 shows that the parameters: the scalar spectral index ($n_s$), the tensor spectral index ($n_t$), the tensor-to-scalar ratio ($r$), the number of e-folding ($N$) are favourably explained by calculations with the obtained potentials. However, the scales of the potentials are off from the corresponding experimental values by about one order of magnitude. It should be mentioned here that our calculations are based on microscopic k-space analysis of  first order cosmological perturbations \cite{Sarkar:2021ird}

\par In conclusion, we have worked in the 6d $T^6/Z_n$ orbifold of $CY_3$, with three intersecting magnetized $D7$ branes, where the superpotentials of two K\"ahler moduli are non-perturbativly corrected in addition to the perturbative corrections \cite{Antoniadis:2019rkh,Becker:2002nn} of the three K\"ahler moduli in the LVS \cite{Balasubramanian:2005zx}. In contrast to our earlier work \cite{Let:2022fmu}, no auxiliary field arises in the present calculations along with the inflaton field, in course of the canonical transformation. Like in the earlier case \cite{Let:2022fmu}, we have obtained here slow-roll inflaton potentials and experimentally-favoured cosmological parameters. However, the scales of inflation do not exactly conform to the experimental ones. Stabilizations of more moduli $\it{e.g.}$, axions might improve the scenario.

\acknowledgments
The authors acknowledge the University Grants Commission for the CAS-II program. AL acknowledges CSIR, the Government of India for NET fellowship. AS and CS acknowledge the Government of West Bengal for granting them Swami Vivekananda fellowship.

\bibliographystyle{eplbib}
\bibliography{biblio}

\begin{thebibliography}{10}
\expandafter\ifx\csname url\endcsname\relax\def\url#1{\texttt{#1}}\fi

\bibitem{Becker:2006dvp}
\Name{Becker K., Becker M. \and Schwarz J.~H.} \Book{{String theory and
  M-theory: A modern introduction}} (Cambridge University Press) 2006.

\bibitem{Antoniadis:2018ngr}
\Name{Antoniadis I., Chen Y. \and Leontaris G.~K.} \REVIEW{Int. J. Mod. Phys.
  A}{34}{2019}{1950042}.

\bibitem{Antoniadis:2020stf}
\Name{Antoniadis I., Lacombe O. \and Leontaris G.~K.} \REVIEW{Eur. Phys. J.
  C}{80}{2020}{1014}.

\bibitem{Conlon:2005jm}
\Name{Conlon J.~P. \and Quevedo F.} \REVIEW{JHEP}{01}{2006}{146}.

\bibitem{Let:2022fmu}
\Name{Let A., Sarkar A., Sarkar C. \and Ghosh B.}
  \REVIEW{EPL}{139}{2022}{59002}.

\bibitem{Agrawal:2018own}
\Name{Agrawal P., Obied G., Steinhardt P.~J. \and Vafa C.} \REVIEW{Phys. Lett.
  B}{784}{2018}{271}.

\bibitem{Palti:2019pca}
\Name{Palti E.} \REVIEW{Fortsch. Phys.}{67}{2019}{1900037}.

\bibitem{Junghans:2018gdb}
\Name{Junghans D.} \REVIEW{JHEP}{03}{2019}{150}.

\bibitem{Akrami:2018ylq}
\Name{Akrami Y., Kallosh R., Linde A. \and Vardanyan V.} \REVIEW{Fortsch.
  Phys.}{67}{2019}{1800075}.

\bibitem{Denef:2018etk}
\Name{Denef F., Hebecker A. \and Wrase T.} \REVIEW{Phys. Rev.
  D}{98}{2018}{086004}.

\bibitem{Conlon:2018eyr}
\Name{Conlon J.~P.} \REVIEW{Int. J. Mod. Phys. A}{33}{2018}{1850178}.

\bibitem{Basiouris:2020jgp}
\Name{Basiouris V. \and Leontaris G.~K.} \REVIEW{Phys. Lett.
  B}{810}{2020}{135809}.

\bibitem{Basiouris:2021sdf}
\Name{Basiouris V. \and Leontaris G.~K.} \REVIEW{Fortsch.
  Phys.}{70}{2022}{2100181}.

\bibitem{Antoniadis:2019doc}
\Name{Antoniadis I., Chen Y. \and Leontaris G.~K.}
  \REVIEW{PoSC}{ORFU2018}{2019}{068}.

\bibitem{Antoniadis:2018hqy}
\Name{Antoniadis I., Chen Y. \and Leontaris G.~K.} \REVIEW{Eur. Phys. J.
  C}{78}{2018}{766}.

\bibitem{Antoniadis:2019rkh}
\Name{Antoniadis I., Chen Y. \and Leontaris G.~K.}
  \REVIEW{JHEP}{01}{2020}{149}.

\bibitem{Kachru:2003aw}
\Name{Kachru S., Kallosh R., Linde A.~D. \and Trivedi S.~P.} \REVIEW{Phys. Rev.
  D}{68}{2003}{046005}.

\bibitem{Haack:2006cy}
\Name{Haack M., Krefl D., Lust D., Van~Proeyen A. \and Zagermann M.}
  \REVIEW{JHEP}{01}{2007}{078}.

\bibitem{Bianchi:2011qh}
\Name{Bianchi M., Collinucci A. \and Martucci L.} \REVIEW{JHEP}{12}{2011}{045}.

\bibitem{Bena:2019mte}
\Name{Bena I., Gra\~na M., Kovensky N. \and Retolaza A.}
  \REVIEW{JHEP}{10}{2019}{200}.

\bibitem{Planck:2018jri}
\Name{Akrami Y. \etal} \REVIEW{Astron. Astrophys.}{641}{2020}{A10}.

\bibitem{Planck:2018vyg}
\Name{Aghanim N. \etal} \REVIEW{Astron. Astrophys.}{641}{2020}{A6} [Erratum:
  Astron.Astrophys. 652, C4 (2021)].

\bibitem{Giddings:2001yu}
\Name{Giddings S.~B., Kachru S. \and Polchinski J.} \REVIEW{Phys. Rev.
  D}{66}{2002}{106006}.

\bibitem{Gukov:1999ya}
\Name{Gukov S., Vafa C. \and Witten E.} \REVIEW{Nucl. Phys. B}{584}{2000}{69}
  [Erratum: Nucl.Phys.B 608, 477--478 (2001)].

\bibitem{Cicoli:2007xp}
\Name{Cicoli M., Conlon J.~P. \and Quevedo F.} \REVIEW{JHEP}{01}{2008}{052}.

\bibitem{AbdusSalam:2020ywo}
\Name{AbdusSalam S., Abel S., Cicoli M., Quevedo F. \and Shukla P.}
  \REVIEW{JHEP}{08}{2020}{047}.

\bibitem{Moritz:2017xto}
\Name{Moritz J., Retolaza A. \and Westphal A.} \REVIEW{Phys. Rev.
  D}{97}{2018}{046010}.

\bibitem{Balasubramanian:2005zx}
\Name{Balasubramanian V., Berglund P., Conlon J.~P. \and Quevedo F.}
  \REVIEW{JHEP}{03}{2005}{007}.

\bibitem{Burgess:2005jx}
\Name{Burgess C.~P., Escoda C. \and Quevedo F.} \REVIEW{JHEP}{06}{2006}{044}.

\bibitem{Becker:2002nn}
\Name{Becker K., Becker M., Haack M. \and Louis J.}
  \REVIEW{JHEP}{06}{2002}{060}.

\bibitem{Green:1999pv}
\Name{Green M.~B. \and Vanhove P.} \REVIEW{Phys. Rev. D}{61}{2000}{104011}.

\bibitem{Green:2010wi}
\Name{Green M.~B., Russo J.~G. \and Vanhove P.} \REVIEW{Phys. Rev.
  D}{81}{2010}{086008}.

\bibitem{Antoniadis:1997eg}
\Name{Antoniadis I., Ferrara S., Minasian R. \and Narain K.~S.} \REVIEW{Nucl.
  Phys. B}{507}{1997}{571}.

\bibitem{Antoniadis:2002tr}
\Name{Antoniadis I., Minasian R. \and Vanhove P.} \REVIEW{Nucl. Phys.
  B}{648}{2003}{69}.

\bibitem{Balasubramanian:2004uy}
\Name{Balasubramanian V. \and Berglund P.} \REVIEW{JHEP}{11}{2004}{085}.

\bibitem{Burgess:2003ic}
\Name{Burgess C.~P., Kallosh R. \and Quevedo F.} \REVIEW{JHEP}{10}{2003}{056}.

\bibitem{Cremades:2007ig}
\Name{Cremades D., Garcia~del Moral M.~P., Quevedo F. \and Suruliz K.}
  \REVIEW{JHEP}{05}{2007}{100}.

\bibitem{Sarkar:2021ird}
\Name{Sarkar A., Sarkar C. \and Ghosh B.} \REVIEW{JCAP}{11}{2021}{029}.

\end{thebibliography}

\end{document}